\documentclass[%
 reprint,
 amsmath,amssymb,
 aps,
 superscriptaddress
]{revtex4-2}

\usepackage{graphicx}
\usepackage{dcolumn}
\usepackage{bm}
\usepackage{comment}
\usepackage{physics}
\usepackage{lipsum}
\usepackage{siunitx}
\usepackage{makecell}
\usepackage{xcolor}
\usepackage[subpreambles=true]{standalone}
\usepackage{import}
\begin{document}

\title{A Terahertz Bandwidth Nonmagnetic Isolator}

\author{Haotian Cheng}
\thanks{These two authors contributed equally.}
\affiliation{Department of Applied Physics, Yale University, New Haven, CT, USA}
\author{Yishu Zhou}
\thanks{These two authors contributed equally.}
\affiliation{Department of Applied Physics, Yale University, New Haven, CT, USA}
\author{Freek Ruesink}
\affiliation{Department of Applied Physics, Yale University, New Haven, CT, USA}
\author{Margaret Pavlovich}
\affiliation{Department of Applied Physics, Yale University, New Haven, CT, USA}
\author{Shai Gertler}
\affiliation{Department of Applied Physics, Yale University, New Haven, CT, USA}
\author{Andrew L. Starbuck}
\affiliation{Microsystems Engineering, Science, and Applications,
Sandia National Laboratories, Albuquerque, NM, USA}
\author{Andrew J. Leenheer}
\affiliation{Microsystems Engineering, Science, and Applications,
Sandia National Laboratories, Albuquerque, NM, USA}
\author{Andrew T. Pomerene}
\affiliation{Microsystems Engineering, Science, and Applications,
Sandia National Laboratories, Albuquerque, NM, USA}
\author{Douglas C. Trotter}
\affiliation{Microsystems Engineering, Science, and Applications,
Sandia National Laboratories, Albuquerque, NM, USA}
\author{Christina Dallo}
\affiliation{Microsystems Engineering, Science, and Applications,
Sandia National Laboratories, Albuquerque, NM, USA}
\author{Matthew Boady}
\affiliation{Microsystems Engineering, Science, and Applications,
Sandia National Laboratories, Albuquerque, NM, USA}
\author{Katherine M. Musick}
\affiliation{Microsystems Engineering, Science, and Applications,
Sandia National Laboratories, Albuquerque, NM, USA}
\author{Michael Gehl}
\affiliation{Microsystems Engineering, Science, and Applications,
Sandia National Laboratories, Albuquerque, NM, USA}
\author{Ashok Kodigala}
\affiliation{Microsystems Engineering, Science, and Applications,
Sandia National Laboratories, Albuquerque, NM, USA}
\author{Matt Eichenfield}
\affiliation{Microsystems Engineering, Science, and Applications,
Sandia National Laboratories, Albuquerque, NM, USA}
\affiliation{James C. Wyant College of Optical Sciences, University of Arizona, Tucson, AZ, USA}
\author{Anthony L. Lentine}
\affiliation{Microsystems Engineering, Science, and Applications,
Sandia National Laboratories, Albuquerque, NM, USA}
\author{Nils T. Otterstrom}
\affiliation{Microsystems Engineering, Science, and Applications,
Sandia National Laboratories, Albuquerque, NM, USA}

\author{Peter T. Rakich}
\affiliation{Department of Applied Physics, Yale University, New Haven, CT, USA}

\date{\today}

\begin{abstract}
Integrated photonics could bring transformative breakthroughs in computing, networking, imaging, sensing, and quantum information processing, enabled by increasingly sophisticated optical functionalities on a photonic chip.
However, wideband optical isolators, which are essential for the robust operation of practically all optical systems, have been challenging to realize in integrated form due to the incompatibility of magnetic media with these circuit technologies.
Here, we present the first-ever demonstration of an integrated non-magnetic optical isolator with terahertz-level optical bandwidth. 
The system is comprised of two acousto-optic frequency-shifting beam splitters which create a non-reciprocal multimode interferometer exhibiting high-contrast, nonreciprocal light transmission. 
We dramatically enhance the isolation bandwidth of this system by precisely dispersion balancing the paths of the interferometer.
Using this approach, we demonstrate integrated nonmagnetic isolators with an optical contrast as high as 28~dB, insertion losses as low as $-$2.16~dB, and optical bandwidths as high as $2$~THz (16~nm). 
We also show that the center frequency and direction of optical isolation are rapidly reconfigurable by tuning the relative phase of the microwave signals used to drive the acousto-optic beam splitters. 
With their CMOS compatibility, wideband operation, low losses, and rapid reconfigurability, such integrated isolators could address a key barrier to the integration of a wide range of photonic functionalities on a chip. 
Looking beyond the current demonstration, this bandwidth-scalable approach to nonmagnetic isolation opens the door to ultrawideband ($>$10~THz) isolators, which are needed to shrink state-of-the-art imaging, sensing, and communications systems into photonic integrated circuits.
\end{abstract}

\maketitle


\section{\label{sec:intro} Introduction}

The ability to generate, manipulate, and detect light using photonic integrated circuits (PICs) of increasing complexity has opened the door to revolutionary advancements in classical~\cite{sun2015single,shen2017deep,feldmann2021parallel,shastri2021photonics} and quantum computation~\cite{sipahigil2016integrated,mehta2020integrated,moody20222022}, optical communications~\cite{roelkens2010iii, shu2022microcomb,rizzo2023massively}, as well as imaging~\cite{riemensberger2020massively,li2023frequency,chen2023breaking} and sensing~\cite{krause2012high,lai2020earth,mohanty2020reconfigurable}. However, the realization of integrated nonreciprocal devices—such as isolators and circulators—has been challenging as they traditionally rely on magneto-optic materials~\cite{bi2011chip,zhang2019monolithic} which are incompatible with complementary metal-oxide-semiconductor (CMOS) technology. Nonreciprocal devices are indispensable for photonic systems because they protect lasers from the destabilizing effects of back-scattered light.  As an alternative to magneto-optical effects, non-reciprocal devices can be implemented by using time modulation~\cite{poulton2012design,sounas2017non,williamson2020integrated}, which has seen impressive demonstrations utilizing electro-optics~\cite{lira2012electrically,yu2022integrated}, optomechanics~\cite{fang2017generalized}, and Kerr nonlinearities~\cite{white2023integrated}.
However, these demonstrations have produced isolation bandwidths that are orders of magnitude smaller than the THz-bandwidth information carrying capacity of optical systems due to constraints associated with the response of integrated resonators~\cite{Sohn2021,Tian2021,herrmann2022mirror,white2023integrated}, the time-modulation frequency~\cite{yu2022integrated}, or the phase-matching bandwidth of traveling wave interaction~\cite{kittlaus2021electrically,zhou2022intermodal}.

Here, we combine nonreciprocal light scattering processes with wideband dispersion engineering to realize a bandwidth-scalable nonmagnetic optical isolator exhibiting $>$1~THz isolation bandwidth. 
The isolator is comprised of two nonreciprocal frequency-shifting beam splitters that form a nonreciprocal multimode interferometer. 
These beam splitters, implemented using electrically-driven acousto-optic scattering, produce a frequency shift through an intermodal (i.e., inter-band) scattering process. 
Within the interferometer, the direction-dependence of the frequency shift produces constructive interference for forward propagating light and cancellation for backward propagating light, yielding isolation contrast as high as 28~dB and insertion losses as low as $-$2.16~dB. 
Precise balancing of the interferometer
dispersion is shown to
radically enhance the isolation bandwidth, yielding isolation bandwidths as high as 2 THz, which we demonstrate using two complementary methods.
Varying the beam splitter phase using a microwave drive, we also demonstrate wideband tuning of the isolator center frequency and re-configurability of the isolator direction. 
Building on these results, rapidly reconfigurable isolators and circulators with much larger bandwidths ($\gg$1~THz) should be achievable, bringing powerful new capabilities to integrated photonics.

\section{\label{sec:concept} System Concept}

\begin{figure*}
    \includegraphics{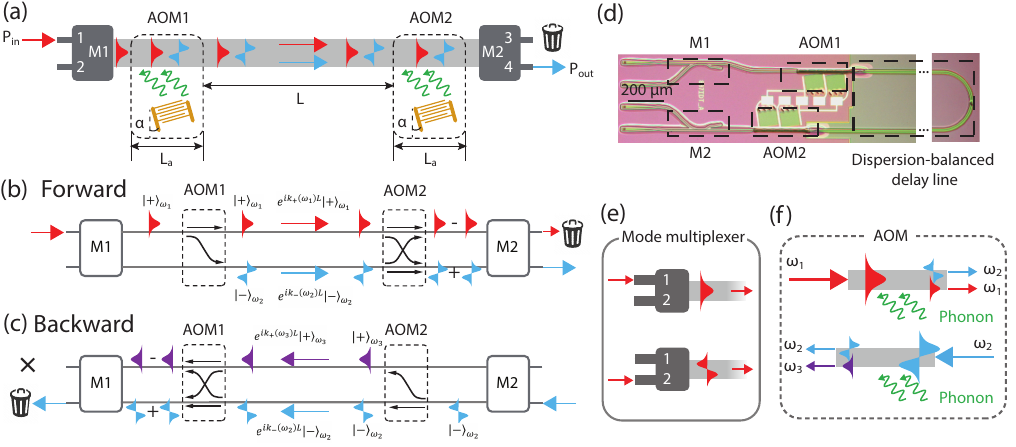}
    \caption{\label{fig1} \textbf{Bandwidth scalable isolator using a dispersion-engineered multimode interferometer. }
    \textbf{a}, Schematic illustration of the nonmagnetic isolator. M1 and M2 represent mode multiplexer 1 and 2, respectively. AOM1 and AOM2 represent acoustic optical modulator 1 and 2, respectively. The angle of the AOMs are designed for phase matching.
    \textbf{b-c}, Dual-rail representation of the isolation mechanism for forward (b) and backward (c) operation.
    \textbf{d}, Optical micrograph of our device, fabricated through a CMOS-foundry process.
    \textbf{e}, Illustration of the spatial mode multiplexer. Each spatial mode is selectively excited by a different input port of the mode multiplexer. 
    \textbf{f}, The nonreciprocal acoustic optical modulator (AOM) in our system partially scatters the forward propagating symmetric mode at frequency $\omega_1$ into an anti-symmetric mode at frequency $\omega_2$, while partially scattering the backward traveling anti-symmetric mode at frequency $\omega_2$ into a symmetric mode at frequency $\omega_3$.}
\end{figure*}

The nonmagnetic isolator (Fig.~\ref{fig1}a) is comprised of a multimode waveguide with two integrated nonreciprocal frequency-shifting beam splitters~\cite{otterstrom2021nonreciprocal} that form a nonreciprocal multimode interferometer. 
Each frequency-shifting beam splitter uses electrically transduced traveling-wave phonons to produce phase-matched intermodal (or inter-band) scattering within a multimode silicon waveguide~\cite{zhou2022intermodal}. 
This acousto-optic interaction transfers energy between a symmetric and antisymmetric TE-like waveguide modes, frequency shifting the scattered waves in the process (see Fig.~1f). 
Two such acousto-optic modulators are cascaded in series to close the paths of the interferometer, producing interference between distinct spatial modes in this multimode waveguide. 
A pair of spatial mode multiplexers (Fig.~\ref{fig1}e) are used to selectively address the symmetric, $|+\rangle$, and antisymmetric, $|-\rangle$, waveguide modes at the input and output of this multimode system. 

We describe the operating principle of this nonmagnetic isolator using a dual-rail representation (Fig.~1b-c), which artificially separates the optical path for each mode. 
In the forward direction (Fig.~1b), light of frequency $\omega_1$ enters Port~1 of mode multiplexer M1 to excite mode $|+\rangle_{\omega_1}$ within the multimode waveguide. 
The notation $|+\rangle_{\omega_1}$ is used to specify the spatial mode, $|+\rangle$, and mode frequency, $\omega_1$, of guided waves in this system. 
As this incident wave traverses the first acousto-optic modulator (AOM~1), a portion of the light is blue-shifted to $\omega_2 = \omega_1+\Omega_o$ and transferred to mode $|-\rangle_{\omega_2}$. 
As these forward-propagating wave components in modes $|+\rangle_{\omega_1}$ and $|-\rangle_{\omega_2}$ propagate over a length ($L$) of the multimode waveguide, they acquire a phase difference ($\phi_\mathrm{f}$) before entering the second acousto-optic modulator (AOM~2).
As these wave components traverse AOM~2, part of the light in mode $|-\rangle_{\omega_2}$ is scattered to $|+\rangle_{\omega_1}$, and part of the light in mode $|+\rangle_{\omega_1}$ is scattered to $|-\rangle_{\omega_2}$, resulting in coherent interference of the transmitted and scattered waves in each spatial mode.
Hence, AOM~1 and AOM~2 can be viewed as frequency-converting beam splitters that produce interference within this multimode interferometer (Fig. 1b).
With the proper choice of passive waveguide length ($L$) and phase difference ($\phi_\mathrm{f}$), wave components scattered by AOM~2 into mode $|+\rangle_{\omega_1}$ destructively interfere; at the same time, the wave components in mode $|-\rangle_{\omega_2}$ constructively interfere, yielding transmission of light from Port~4 of mode multiplexer M2.

Due to the nonreciprocal nature of inter-modal acousto-optic scattering, waves entering Port~4 do not retrace the same path in the backward direction. 
A reflected wave of frequency $\omega_2$ that enters AOM~2 through mode $|-\rangle_{\omega_2}$, is blue-shifted to frequency $\omega_3 = \omega_2+\Omega_o$ and scattered to mode $|+\rangle_{\omega_3}$. 
In contrast to the forward-propagating case, the light in mode $|+\rangle_{\omega_3}$ is higher in frequency than that in mode $|-\rangle_{\omega_2}$.
Hence, wave components $|+\rangle_{\omega_3}$ and $|-\rangle_{\omega_2}$ accumulate a distinct relative phase ($\phi_\mathrm{b}$) as they propagate through the passive waveguide section in the backward direction; consequently, they do not interfere the same way when recombined, giving rise to nonreciprocal light transmission. 
Optical isolation is produced when paths of this multimode system are designed to simultaneously produce constructive interference in the forward direction and destructive interference in the backward direction; we will see that these criteria require that the nonreciprocal phase difference, $\Delta\phi$, satisfies the conditions $\Delta\phi =\phi_\mathrm{f}-\phi_\mathrm{b} = \pi$ and $\phi_\mathrm{f}=0$. 
We will show that dispersion engineering strategies can be used to satisfy this condition over a large bandwidth, yielding high-contrast and wideband optical isolation. 

Independent control of the relative phases in each direction, $\phi_\mathrm{f}$ and $\phi_\mathrm{b}$, is critical for achieving high-contrast isolation, thereby setting strict design requirements on the system. 
In the forward direction, incident light at frequency $\omega_1$ leads to relative phase between two optical paths $\phi_\mathrm{f} (\omega_1) = k_+ (\omega_1) L - k_- (\omega_2) L $, where $k_+(\omega)$  and $k_-(\omega)$ are the optical wave vectors at frequency $\omega$ of the symmetric and anti-symmetric mode, respectively. 
However, injecting the transmitted signal at frequency $\omega_2$ back into the system, the relative phase becomes $\phi_\mathrm{b} (\omega_2) = k_+ (\omega_3) L - k_- (\omega_2) L = \phi_\mathrm{f} + 2v_\mathrm{g+} \Omega_0 L $, due to the distinct frequencies in symmetric mode relative to forward direction. 
Here, $v_\mathrm{g+}$ is the optical group velocity of the symmetric mode. 
Consequently, the nonreciprocal phase difference is $\Delta \phi = 2v_\mathrm{g+} \Omega_0 L$, and we require $\Delta \phi = \pi $ to achieve isolation. 
In this system, the optical path length ($L$) is chosen to satisfy this condition. 

The isolation bandwidth of our system is determined by the difference in effective group delay between the paths of the interferometer. Imbalanced path yields fringes in transmission spectrum. 
Since the fringe period of an interferometer is determined by the differential group delay of the optical paths, the fringe period can be greatly extended by balancing the group delays of the interferometer, increasing the available bandwidth of high-contrast isolation. 
Hence, to achieve wideband operation, it is essential to minimize the difference in group delay between two paths of the nonreciprocal multimode interferometer to maintain a constant phase difference as the operation frequency varies.
For example, in the case of the forward propagation, a shift in the incident optical frequency, $\Delta\omega$, produces the relative phase of $\phi_\mathrm{f} (\omega_1 + \Delta \omega) \cong \phi_\mathrm{f} (\omega_1) + (v_\mathrm{g+} L-v_\mathrm{g-} L) \Delta \omega$. 
Hence, by satisfying the condition, $v_\mathrm{g+} = v_\mathrm{g-}$, the frequency-dependent phase shift vanishes to first order, enabling wideband isolation. 
This is accomplished by finely tuning the dispersion properties of the optical paths to balance the group delays that accumulate in the two different paths. 

In what follows, we introduce two distinct methods for wideband dispersion engineering. 
The first method utilizes a two-ridge waveguide, where the geometry of the separate ridges are individually tuned to control the group velocities of the two optical modes, such that $v_\mathrm{g+}  \cong v_\mathrm{g-} $. 
The second method features a dual-waveguide design, dividing the two optical modes into separate waveguides.
In this case, the relative phase between the two optical paths becomes $\phi_\mathrm{f} (\omega_1 + \Delta \omega) \cong \phi_\mathrm{f} (\omega_1) + (v_\mathrm{g+} L_{+}-v_\mathrm{g-} L_{-}) \Delta \omega$, where $L_+$ and $L_-$ are the lengths of the optical paths associated with the $|+\rangle$ and $|-\rangle$ modes. 
By independently adjusting the waveguide lengths $L_+$ and $L_-$ (Section~III.B), we effectively balance their group delays over a wide spectral range. 

\begin{figure*}
    \includegraphics{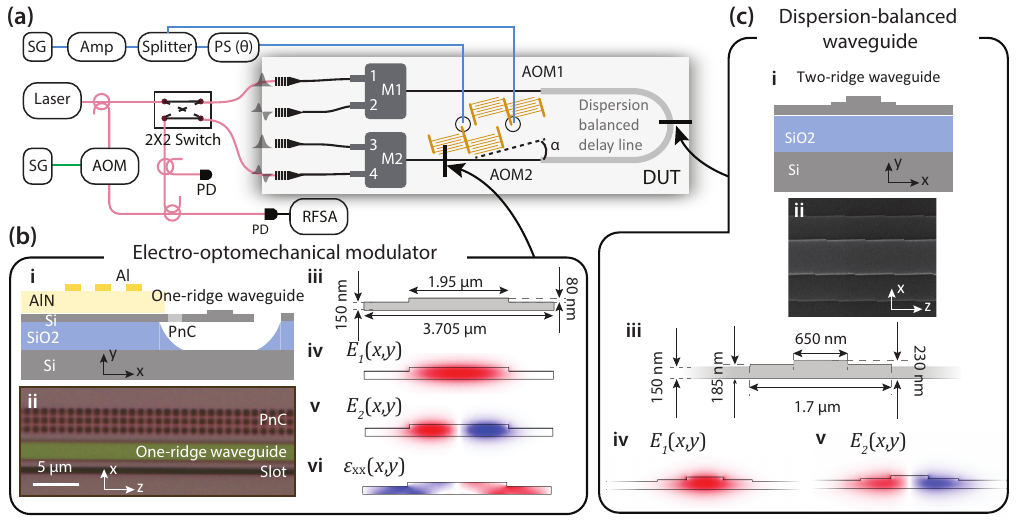}
    \caption{\label{fig2} \textbf{Experimental setup. }
    \textbf{a}, Experimental setup for characterizing the forward and backward operation of our isolator.
    SG: signal generator; Amp: RF amplifier; Splitter: 50/50 power splitter; PS: RF phase shifter; AOM: acousto-optic modulator; PD: photodiode; RFSA: RF spectral analyzer. Pink lines represent optical connections, blue lines represent high frequency ($\sim$2GHz) electrical connections and green lines represent low frequency (55MHz) electrical connections.
    \textbf{b}, The on-chip acousto-optic modulator (AOM) comprises a piezoelectric interdigital transducer (IDT) fabricated on AlN and an optomechanical waveguide  crafted from a suspended single silicon ridge, with crossection illustrated in \textbf{i}. 
    \textbf{ii} and \textbf{iii} present an optical micrograph and the designed dimensions of the optomechanical waveguide, which supports two optical modes (electric field profile shown in \textbf{iv}, \textbf{v}) and a phonon mode (strain field shown in \textbf{vi}). 
    \textbf{c}, The dispersion-balanced waveguide is an unsuspended two-ridge waveguide, with crossection illustrated in \textbf{i}. 
    \textbf{ii} and \textbf{iii} show a SEM image and the designed dimensions of the two-ridge waveguide, which tunes the dispersion properties of the symmetric (\textbf{iv}) and antisymmetric (\textbf{v}) modes separately through the designs of the narrow and wide silicon ridges. 
    }
\end{figure*}

\begin{figure*}
    \includegraphics{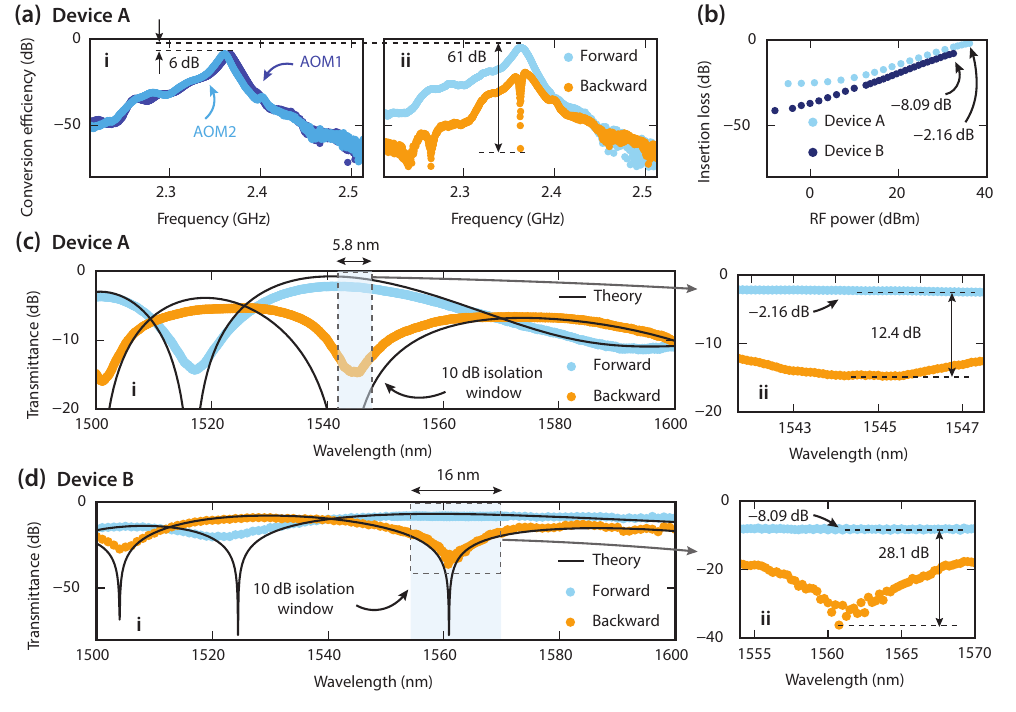}
    \caption{\label{fig3} \textbf{Wideband isolator based on a dispersion-engineered waveguide. }
    \textbf{a}, We measure the mode conversion efficiency as a function of the microwave drive frequency while driving only AOM 1 (dark blue in \textbf{i}) and driving only AOM 2 (light blue in \textbf{i}), which demonstrates almost identical responses. 
    When both AOMs are simultaneously driven (\textbf{i}), the forward (backward) operation of our device equates to the sum (difference) of the responses from the two AOMs, leading to the high isolation contrast exhibited by our device. 
    \textbf{b}, The insertion loss decreases as we increase the microwave drive power, approaching $-$2 dB. The maximum RF power is limited by the power handling of the device.
    \textbf{c}, The transmission of Device A operating in forward (blue) and backward (orange) directions, which demonstrates the 10~dB isolation bandwidth of 5.8~nm (725~GHz, shaded blue window in \textbf{i}, enlarged in \textbf{ii}). Over the isolation bandwidth, the device has $-$2.16 dB insertion loss and 12.4 dB peak isolation ratio. 
    Measurements deviate slightly from theoretical predictions (black) due to pump depletion and intermodal crosstalk in the passive photonic components of the device. Note that higher contrast is seen between forward and backward transmission in (\textbf{a}) because these frequency-resolved measurements display only the frequency-shifted components of the transmitted light field.
    \textbf{d}, The transmission of Device B operating in forward (blue) and backward (orange) directions, which demonstrates the 10~dB isolation bandwidth of 16~nm (2~THz, shaded blue window in \textbf{i}, enlarged in \textbf{ii}). Over the isolation bandwidth, the device has $-$8.09 dB insertion loss and 28.1 dB peak isolation ratio.
    }
\end{figure*}

\section{\label{sec:results} Experimental Results}

Next, we describe nonmagnetic isolator designs that utilize two complementary dispersion-balancing techniques; both were fabricated in silicon using a CMOS-compatible process.
The AOMs used to construct nonreciprocal interferometers within both of these systems utilize a membrane-suspended optomechanical waveguide (Fig. \ref{fig2}b) that is designed to produce efficient intermodal acousto-optic scattering. 
Elastic waves supplied by an angled interdigitated transducer (IDT) (Fig. \ref{fig2}a) excite guided phonons of the appropriate wave vector and frequency to facilitate efficient scattering between distinct spatial modes supported by the optical waveguide. 

The optomechanical waveguide seen in Fig. \ref{fig2}b produces simultaneous guidance of both optical and acoustic waves, necessary to maximize acousto-optic interaction strength.
Optical waves are confined through total internal reflection within the central ridge (Fig.~\ref{fig2}b.i) while acoustic waves are confined by a phononic crystal and an air slot on either side of the ridge (Fig.~\ref{fig2}b.ii). 
This optomechanical waveguide supports symmetric ($|+\rangle$) and antisymmetric ($|-\rangle$) waveguide modes (Fig.~\ref{fig2}b.iv and v) as well as a high-\textit{Q} guided acoustic mode (Fig.~\ref{fig2}b.vi), with waveguide dimensions (Fig.~\ref{fig2}b.iii) chosen to maximize the acousto-optic scattering efficiency. 
The IDT, which is patterned on a piezoelectric AlN layer that sits atop the silicon layer (Fig.~\ref{fig2}b.i), emits $\sim$2.36~GHz phonons that resonantly tunnel into the waveguide through the partially-transmitting phononic crystal patterned in the silicon layer. 
The IDT pitch and angle ($\alpha$) are chosen to produce guided phonons of the specific frequency and wave vector needed to facilitate phase-matched acousto-optic intermodal scattering between the optical modes within the ridge waveguide (for AOM details, see Ref.~\cite{zhou2022intermodal}).

\subsection{Wideband optical isolation with a dispersion balanced multimode interferometer}

We begin by examining the performance of an isolator (Fig. \ref{fig3}) that utilizes a dispersion-engineered multimode waveguide to satisfy the conditions for wideband nonreciprocity described in Section \ref{sec:concept}.
In this system, the group delay experienced by the symmetric ($|+\rangle$) and antisymmetric ($|-\rangle$) waveguide modes are precisely balanced by optimizing the dimensions of dual-ridge multimode waveguide seen in Fig.~\ref{fig2}c.i and ii. 
The narrow and wide silicon ridges control the dispersion of the symmetric (Fig.~\ref{fig2}c.iv) and antisymmetric (Fig.~\ref{fig2}c.v) modes, respectively; the chosen waveguide dimensions (see Fig.~\ref{fig2}c.iii) equalize of the group delay of the $|+\rangle$ and $|-\rangle$ modes, which define the two paths of the interferometer (see in Supplementary Information Sec.II (B-C) for further details). 
To meet the necessary conditions for constructive (destructive) interference in the forward (backward) direction, a dispersion-engineered waveguide of 7.64~mm length is used to link AOM~1 and AOM~2 within the interferometer.

The performance of the AOMs within the isolator are analyzed using the experimental setup of Fig.~\ref{fig2}a (see Supplementary Information Sec.~V for details).
The scattering efficiency and spectral response of each AOM is analyzed by mixing the transmitted optical waves with an electrically synthesized optical local oscillator (LO) while measuring the detected heterodyne spectrum.  
The measured RF-frequency-dependent conversion efficiencies of AOM 1 and AOM 2 are shown in Fig.~\ref{fig3}a.i at a constant microwave drive power of 32.39~dBm. The conversion efficiency is defined as the fraction of incident laser power entering port 1 that is blue-shifted before exiting the system through port 4.
Measurements reveal a peak acousto-optic scattering efficiency of -4.6~dB at a  $\sim$2.36~GHz drive frequency; this peak in conversion efficiency identifies the frequency at which phonons emitted by the IDT excite the guided acoustic wave resonance, having a Q-factor of $\sim230$.

When both AOM~1 and AOM~2 are electrically driven, transmission measurements reveal constructive (destructive) interference of the acousto-optically scattered wave components in the forward (backward) direction, producing high-contrast optical isolation. 
To analyze the response of the system, we again perform frequency-resolved transmission measurements (Fig.~\ref{fig3}a.ii), which show the portion of incident laser power that is blue-shifted by the AOMs as it traverses the system.
To examine the forward operation of the isolator, we inject 1547.5~nm light into port~1 and measure the frequency-shifted optical transmission from port~4 while sweeping the drive frequency of AOM~1 and AOM~2. 
Transmission measurements in the forward direction (blue) reveal a 6 dB increase in light transmission relative to that of a single AOM (Fig.~\ref{fig3}a.i) due to coherent summation of the scattered wave components from both AOMs. 
To examine the backward operation of the isolator, we inject 1547.5~nm light into port 4 and measure the optical transmission from port 1. 
For the case of backward transmission, the phase-coherent addition of scattered wave components from both AOMs produces destructive interference, corresponding to 61~dB of nonreciprocal contrast at 2.36~GHz drive frequencies.

To highlight the potential for both low optical losses and wide isolation bandwidths using this technique, we examine the performance of two devices with the same design, Device~A and Device~B, having complementary performance characteristics. 
Device~A was released using liquid-phase HF (l-HF) and has higher acousto-optic scattering efficiencies.
Since the insertion loss of the isolator is directly determined by acousto-optic conversion efficiency, measurements of isolator insertion loss as a function of microwave drive power (Fig.~\ref{fig3}b) reveal lower insertion losses at higher microwave drive intensities for both devices.
Note that the finite isolator transmission at low microwave drive powers primarily results from intermodal crosstalk in the passive circuit components (see Supplementary Information Sec.III B).
As seen from  Fig.~\ref{fig3}b, the higher acousto-optic scattering efficiencies of Device~A translate to a lower insertion loss ($-$2.16 dB) for forward-transmitted light at the maximum microwave drive power.
Device~B, which was released using vapor-phase HF (v-HF), exhibits lower acousto-optic scattering efficiencies, translating to higher ($-$8.09 dB) isolator insertion losses (see Fig.~\ref{fig3}b). 
However, lower inter-modal crosstalk within the passive circuit elements of Device~B translates to higher isolation contrast, and the more optimal waveguide dimensions of Device B also produced improved dispersion balancing, which contributes to wider isolation bandwidths (See Supplementary Information Sec. II B).
The distinct AOM performance characteristics of Devices A and B result from the different release methods used (See Supplementary Information Sec. I).

\begin{figure*}
    \includegraphics{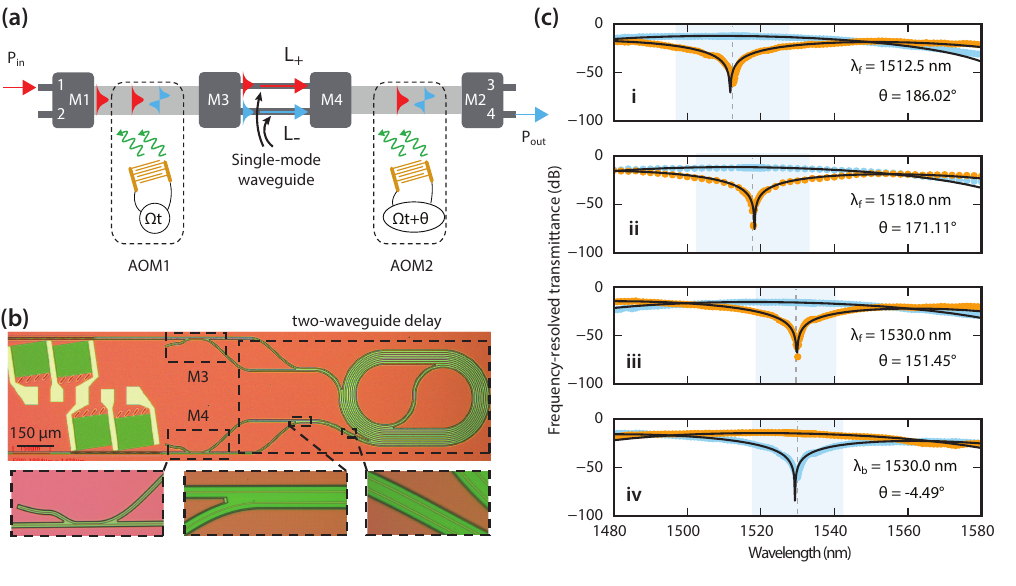}
    \caption{\label{fig4} \textbf{Wideband isolator based on a dispersion-balanced two-waveguide delay. }
    \textbf{a}, The second dispersion engineering method incorporates two extra mode multiplexers to split the two waveguide modes into two single-mode waveguides. 
    \textbf{b}, An optical micrograph of the fabricated device. 
    The two-waveguide delay line is configured in a spiral shape, taking advantage of its robust dispersion properties that can withstand waveguide bends without significant degradation. 
    \textbf{c}, Heterodyne measurement results that filter out intermodal crosstalk. Our device exhibits wavelength tunability (\textbf{i-iii}) and reconfigurability of the isolation direction (\textbf{iii} and \textbf{iv}), which can be adjusted by controlling the relative phase difference $\theta$ between the microwave drives for AOM 1 and AOM 2. 
    }
\end{figure*}
Next, we analyze the wideband performance of these isolators by sweeping the laser wavelength and measuring the total optical power transmittance while the AOMs are driven at their peak frequencies and maximum RF powers.
Transmission measurements in both the forward (orange) and backward (blue) directions, reveal high-contrast isolation for Devices~A (Fig.~\ref{fig3}c) and B (Fig.~\ref{fig3}d). 
These measurements were performed using direct optical power detection, such that total optical power exiting Port 4 (Port 1) in the forward (backward) direction is compared with the optical power entering Port 1 (Port 4); no spectral filtering is being performed. 
Device A (Device B) exhibits a peak isolation contrast of 12.4 dB (28.1 dB) and an insertion loss of $-$2.16 dB ($-$8.09 dB) at 1547.5 nm (1560.7 nm), corresponding to the center of the isolation bandwidth;
these spectral measurements reveal a 10 dB isolation bandwidth of 5.8 nm (16 nm) corresponding to a frequency bandwidth of 0.73 THz (2 THz), representing orders of magnitude larger isolation bandwidth than previous nonmagnetic isolators (see supplement Section IV).

This broad isolation bandwidth is enabled by precisely balancing the group velocities of the $|+\rangle$ and $|-\rangle$ modes in the dual-ridge waveguide design.
The observed fringe period is consistent with a group delay difference of $\sim\!50$~fs ($\sim\!30$~fs) within Device~A (Device~B); observed variations in the fringe period arise from the distinct group velocity dispersion (GVD) of the two modes within the multimode waveguide (Supplementary Information Sec.II B-C).
The theoretical isolator performance (black) is based on an analytical model (see Supplementary Information Sec. II A) and shows good agreement with the measured isolator response.
Interestingly, this isolation bandwidth can made even larger with more precise balancing of the dispersion in each arm of the interferometer.

\subsection{Tunable wideband nonreciprocity with a dual waveguide interferometer}

Next, we present a complementary strategy for wideband dispersion engineering, and we use it to demonstrate frequency tunability and rapid reconfigurability of the nonreciprocal response. 
Through this second approach, in place of a dispersion-engineered multimode waveguide, we instead use two separate single-mode waveguides (Fig.~\ref{fig4}a) to precisely match the group delay in each arm, as required to achieve wideband nonreciprocity.   
As seen in Fig.~\ref{fig4}, waves propagating in modes $|+\rangle$ and $|-\rangle$ at the output of AOM~1 are demultiplexed by M3 and transmitted through two identical single-mode waveguides before being recombined by M4 and entering AOM~2.
Since both arms of the interferometer are identical, this approach produces a constant group-delay difference over all wavelengths, meaning that the fringe period of the interferometer does not change as a function of wavelength. 

We examine the nonreciprocal response of the micro-fabricated device seen in Fig. \ref{fig4}b. This system exhibits lower acousto-optic scattering efficiencies ($-12.29$ dB) due to reduced electromechanical couplings with none optimal IDT design in this device.
Since the acousto-optic scattering efficiency is not appreciably larger than the intermodal crosstalk, it is not possible to demonstrate optical isolation using direct power detection, as we did in Fig.~\ref{fig3} for the design based on dispersion-engineered multimode waveguide.
Instead, we use optical heterodyne measurements to analyze the constructive and destructive interference between the acousto-optically scattered tones and illustrate the the unique advantages of this dispersion-balancing approach.
Figure \ref{fig4}c.i shows the frequency-resolved transmission measurements in forward and backward directions while both AOM~1 and AOM~2 are electrically driven. 
In the forward direction (blue), the frequency-shifted components of the transmitted wave produce constructive interference, with maximum of conversion efficiency at 1515.4 nm. 
In the backward direction (orange), we observe cancellation of the frequency-shifted wave components, corresponding to 48.7 dB of nonreciprocal contrast at 1512.5 nm. 
The wavelength-dependent transmission in both forward and backward directions show excellent agreement with the predicted nonreciprocal response (black), revealing a 10~dB nonreciprocal bandwidth of 31~nm, highlighted in Fig. \ref{fig4}c.i.
Note that the nonreciprocal contrast is much larger than those of Fig.~\ref{fig3} since these frequency-resolved transmission measurements artificially eliminate inter-modal crosstalk associated with the passive components within the system. 

This wide nonreciprocal bandwidth is enabled by the precise balancing group delay produced in each arm of the interferometer; the observed fringe period of 83~nm corresponds to a 31-femtosecond difference in group delay between the optical paths. 
To achieve this, it was necessary to use slightly different waveguide lengths ($L_+-L_-=10.6$ \textmu m) in upper and lower arms of the interferometer to compensate for dissimilarities in the group delay experienced by modes $|+\rangle$ and $|-\rangle$ as they traverse the acousto-optic modulators and mode multiplexers (see supplement III for details).
In principle, the nonreciprocal bandwidths of this system can be dramatically enhanced by more precisely adjusting the relative lengths of the upper and lower waveguides within the interferometer to improve further the group delay balancing in the system.  

Furthermore, the nonreciprocal passband and direction are readily reconfigured by controlling the relative phase of the microwave tones driving AOM~1 and AOM~2. 
Since the acousto-optically scattered wave components inherit the phase of the microwave drive tone, we can shift the interferometer fringes as well as the passband center frequency by tuning the relative phase of the microwave tones that are used to drive AOM~1 and AOM~2. 
Figures \ref{fig4}c.i-iii show a sequence of frequency-resolved transmission measurements as the microwave phase ($\theta$) at the input of AOM~1 is varied by 35 degrees, tuning the nonreciprocal passband by 18~nm.
Moreover, by implementing a 180-degree change in the phase of the microwave drive, as seen in Fig. \ref{fig4}c.iv, we see that the direction of nonreciprocal light transmission becomes reversed, enabling rapid reconfigurability of the direction of isolation.

These two methods for extending the isolator bandwidth--- using a single dispersion-engineered multimode waveguide or two identical waveguides--- have distinct advantages. 
While we have demonstrated low losses and wideband optical isolation using the dispersion-engineered multimode waveguide, control of the waveguide geometry with nanometer-scale precision is necessary, which is not always practical. We have shown that comparable nonreciprocal bandwidths can be achieved using two identical waveguides to ensure that the group delay produced in each arm is well-matched. 
This approach eliminates the need for tight dimensional control of the waveguide geometry; both waveguides need only have identical dimensions, a condition easily met with most fabrication processes. 
However, this second approach places more stringent requirements on mode multiplexer performance to achieve high contrast isolation. 
Nevertheless, with further design optimization, either approach can be used to produce high-performance wideband isolators in any number of photonic platforms. 


\section{\label{sec:disc} Discussion and conclusion}

In conclusion, we have demonstrated a new approach for wideband nonmagnetic optical isolation that combines nonreciprocal light scattering processes with dispersion engineering to realize a bandwidth-scalable nonmagnetic isolator.
Using this approach, we have demonstrated high isolation contrast over bandwidths as high as 2 THz, representing a major advance for nonmagnetic isolators in integrated photonics. These results correspond to a 30-fold increase in bandwidth over waveguide-based devices \cite{kittlaus2021electrically,zhou2022intermodal} and a over 1000-fold increase relative to resonator-based devices ~\cite{yang2020inverse,Sohn2021,Tian2021,white2023integrated}.
Moreover,  with more precise dispersion balancing, isolation bandwidths exceeding 10 THz can be achieved, extending the capabilities of such integrated nonreciprocal technologies.


Further improvements in the AOM performance will translate to lower isolator insertion losses and greatly reduced power consumption, which is necessary for practical integrated technologies. 
The insertion losses of these isolators are currently limited by the AOM scattering efficiency.
Hence, a modest increase in AOM scattering efficiency would eliminate this source of excess loss, reducing the insertion losses to $\sim\!0.8$~dB levels (limited by waveguide loss). 
Additionally, with further refinement of the electromechanical transducers that drive the AOMs, the microwave power necessary to power this isolator can be reduced by several orders of magnitude.
Simulations reveal that $\sim 400$ \textmu W of acoustic power guided within the optomechanical waveguide is sufficient to produce complete acousto-optic conversion within the AOMs.
Hence, with further optimization of the electromechanical transducer design it should be possible to reduce reduce the power consumption of the isolator to milliwatt levels.  

Looking beyond this demonstration, these concepts for creating and controlling wideband nonreciprocity are broadly applicable to other photonic platforms. 
While we used acousto-optic scattering to make this demonstration, other intermodal scattering mechanism, such as electric-optical modulation\cite{lira2012electrically}, can in principle be used to achieve the same functionality, opening the door to wideband isolation using a broader range of interactions in a variety of materials platforms. 
Rapid electrical reconfigurability and tunabilty of the isolator passband also offer intriuging possibilities for extending the bandwidth and functionalities of photonic systems, with potential impacts in sensing, communications, computing, and frequency metrology.

\bibliography{MainRef}

\noindent \textbf{Corresponding authors}: \href{mailto:peter.rakich@yale.edu}{peter.rakich@yale.edu} and \href{mailto:yishu.zhou@yale.edu}{yishu.zhou@yale.edu}\vspace{6pt}

\noindent \textbf{Funding}: 
This research was developed with funding from the Defense Advanced Research Projects Agency (DARPA LUMOS) under award No. HR0011048577 and the National Science Foundation (NSF) under Award No. 2137740. 
The views, opinions and/or findings expressed are those of the author and should not be interpreted as representing the official views or policies of the Department of Defense, National Science Foundation, or the U.S. Government.
Distribution Statement A - Approved for Public Release, Distribution Unlimited. 
This material is based upon work supported by the Laboratory Directed Research and Development program at Sandia National Laboratories. 
Sandia National Laboratories is a multi-program laboratory managed and operated by National Technology and Engineering Solutions of Sandia, LLC., a wholly owned subsidiary of Honeywell International, Inc., for the U.S. Department of Energy's National Nuclear Security Administration under contract DE-NA-0003525. 
This paper describes objective technical results and analysis. 
The views, opinions, and/or findings expressed are those of the authors and should not be interpreted as representing the official views or policies of the U.S. Department of Energy, U.S. Department of Defense, or the U.S. Government.

\noindent \textbf{Author Contributions}:

H.C., Y.Z., N.T.O, A.L.L., and P.T.R. led the project and conceived of the device physics and experiment.

H.C., Y.Z. designed and measured dispersion engineered waveguide devices.

Y.Z., H.C. designed and measured dual waveguide devices.

F.R., M.P. and S.G. contributed to AOM design.

A.L.S., A.J.L., A.T.P., D.C.T., C.D., M.B., K.M.M., M.G., A.K., A.L.L., M.E., and N.T.O. fabricated the devices.

Y.Z., H.C., P.T.R. wrote the paper with input from all authors.

All authors contributed to the design and discussion of the results.

\clearpage
\newpage
\onecolumngrid

\clearpage
\newpage
\begin{center}
    \begin{Large} 
    Supplementary Information: A Terahertz Bandwidth Nonmagnetic Isolator
    \end{Large}
\end{center}
\renewcommand{\theequation}{S.\arabic{equation}}
\renewcommand{\thefigure}{S\arabic{figure}}

\section{Electro-optomechanical scattering}
\subsection{Phase matching condition}
In this section, we address an incident acoustic phonon characterized by a frequency of $\Omega/2\pi$ and associated wavevector along the waveguide direction, ${q(\Omega)}$. In order to achieve effective forward operation of the isolator via acoustic optical scattering, it is crucial that the condition for phase matching is met. This can be mathematically expressed as:

\begin{equation}
k_\mathrm{+}(\omega_0)+q(\Omega)=k_\mathrm{-}(\omega_0+\Omega)
\end{equation}

\begin{figure*}
    \includegraphics{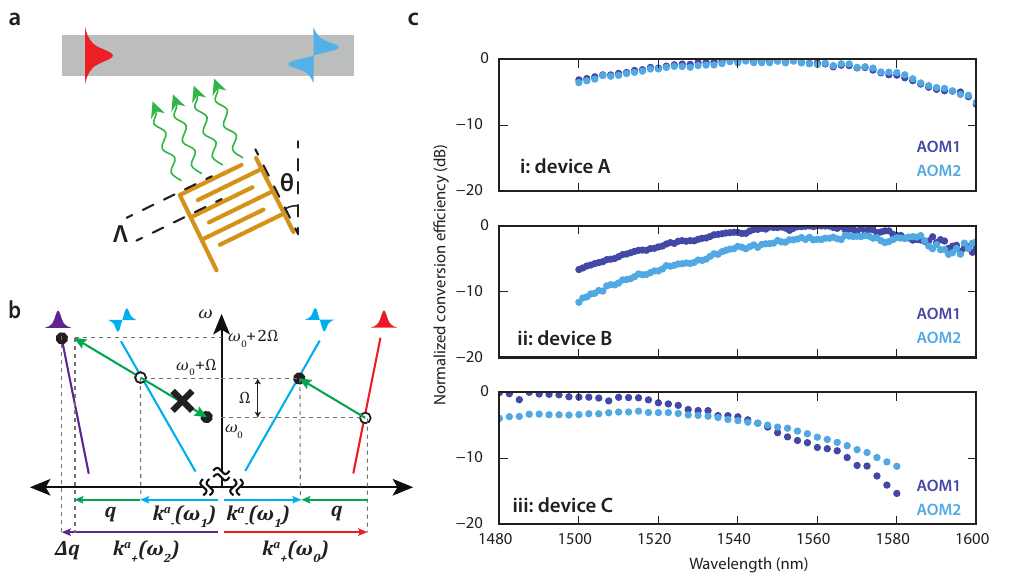}
    \caption{\label{figS1}
    \textbf{a}, Schematic of single section electro-optomechanical scattering. 
    \textbf{b}, Dispersion diagram of optomechanical scattering.
    \textbf{c}, Single section measurement of three different devices presented in the main text. }
\end{figure*}

In the above equation, $k_+(\omega_0)$ and $\omega_0/2\pi$ represent the wavevector and frequency of the incident light in the symmetric mode, respectively, while $k_-(\omega_0+\Omega)$ is the wavevector of the anti-Stokes scattered light in the anti-symmetric mode, as they are shown in Fig.~\ref{figS1}(b).

If we examine a scenario where the incident optical frequency is slightly off the ideal phase matching condition, a phase mismatch will inevitably occur. This phase mismatch can be quantitatively evaluated using the following equation:

\begin{equation}
\Delta q_\mathrm{pm}(\omega)= k_+(\omega)+q(\Omega)-k_-(\omega+\Omega)=\frac{n_\mathrm{g,+}-n_\mathrm{g,-}}{c}(\omega-\omega_0)
\end{equation}

In this equation, $n_\mathrm{g,1(2)}$ signifies the group index of the symmetric (anti-symmetric) modes within the active waveguide.

Following the establishment of the phase mismatch, the frequency response of modulation will bear a proportional relationship to $|\int_0^{L_\mathrm{a}} \exp{i\Delta q_\mathrm{pm}z}dz|^2 = L_\mathrm{a}^2\mathrm{sinc^2} (\Delta q_\mathrm{pm}L_\mathrm{a}/2)$, with $L_\mathrm{a}$ being the interaction length within the active region. Given our interaction length $L_\mathrm{a}$ is relatively short, $\Delta q_\mathrm{pm}L_\mathrm{a}$ will be small over a broadband of optical wavelength. And even in backward direction, we have a phase mismatch term $\Delta q_\mathrm{pm} = 2\frac{n_\mathrm{g,+}}{c}\Omega$, as is shown in Fig.~\ref{figS1}(b), the anti-Stokes scattering can still happen since $\Delta q_\mathrm{pm} L_\mathrm{a}$ is small. However the Stokes process won't happen in backward operation because of phase mismatch being too large. Such phase matching condition provides us a non-reciprocal acoustic optical modulation which is essential to make an isolator.

It is important to note that the acoustic wavevector is determined by the IDT tooth period ($\Lambda$) and angle ($\theta$), as given by:

\begin{equation}
q(\Omega)=\frac{2\pi}{\Lambda}\sin{\theta}
\end{equation}

Therefore, based on equation S.1, the center frequency of conversion will be influenced by $\Lambda,\theta$ parameters of the IDT.

\begin{figure*}
    \includegraphics{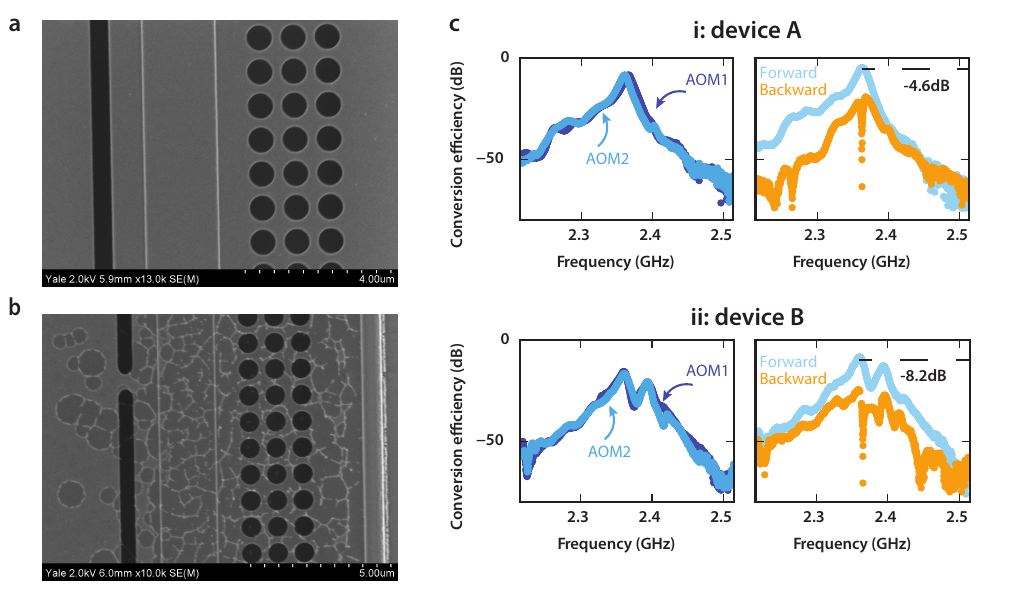}
    \caption{\label{figS2}
    \textbf{a, b}, SEM image of optomechanical scattering active region suspended by liquid HF and vapor HF respectively.
    \textbf{c}, RF frequency response of electro-optomechanical section with 32dBm RF drive power. Device A and device B are released by liquid HF and vapor HF respectively. }
\end{figure*}

Here our IDT tooth period is fixed at 3.5$\mathrm{\mu m}$, and IDT angle is different for three devices we showed in the main text. The single section conversion efficiency characterization is shown in Fig.~\ref{figS1}(c). Device A, B and C have the IDT angle of 8.59\textdegree, 8.77\textdegree and 7.94\textdegree, showing different center wavelength and broadband response. Notice device A is much better in terms of uniformity for two AOMs. This is because device A is released by liquid HF while device B and C are released by vapor HF. We observed vapor HF will leave residues on suspended region, as is shown in Fig.~\ref{figS2}(a-b), that will lower the scattering efficiency by 3.6dB (Fig.~\ref{figS2}(c)) and increase the variations between AOMs and devices.

The microwave power efficiency for electro-optomechanical scattering can be further optimized by several simple design optimization.
First we can suspend the IDT such that most of the acoustic energy is confined in Si layer rather than in the bottom silica cladding of the device layer\cite{kittlaus2021electrically}. Secondly, we can improve the impedance matching of IDT to 50~$\Omega$ by increasing the IDT area. It will also improve the acoustic optical coupling for longer interaction length.
\subsection{Transfer matrix analysis in dual-rail representation}

In this section, we present a theoretical analysis of our nonreciprocal multimode interferometer, presenting a way to explain its operation principles. 
For forward operation, we describe the system's response as follows
\begin{equation}
    \begin{bmatrix} |+\rangle_{\mathrm{o},\omega_1} \\ |-\rangle_{\mathrm{o},\omega_2} \end{bmatrix}  
    = t_\mathrm{a}^f(\phi_\mathrm{RF2}) \cdot t_\mathrm{delay} \cdot t_\mathrm{a}^f(\phi_\mathrm{RF1}) 
    \begin{bmatrix} |+\rangle_{\mathrm{i},\omega_1}\\ |-\rangle_{\mathrm{i},\omega_2} \end{bmatrix}.
\end{equation}
Here, the transfer matrices $t^f_\mathrm{a} (\phi_\mathrm{RF1(2)})$ represent the first and second AOM section with input RF phase $\phi_\mathrm{RF1(2)}$, respectively, while $t_\mathrm{delay}$ accounts for the delay introduced between these AOMs.

With waveguide losses disregarded, the transfer matrix for the AOM section can be expressed as \cite{zhou2022intermodal}
\begin{multline}
    t_{a}^f (\phi_\mathrm{RF}) = 
    \begin{bmatrix}
        e^{ \left[ i\left( k_+- \frac{\Delta q_\mathrm{pm}}{2} \right) L_\mathrm{a} \right]} 
        \left\{ \cos{\left[ \beta L_\mathrm{a} \right] } + \frac{i \Delta q_\mathrm{pm}}{2 \beta}  \sin{\left[ \beta L_\mathrm{a} \right] } \right\} &
        - e^{ \left[ i\left( k_+- \frac{\Delta q_\mathrm{pm}}{2} \right) L_\mathrm{a} \right]}  \frac{i g^* \bar{b}^*}{ \sqrt{v_1v_2} } \sin{\left[ \beta L_\mathrm{a} \right] } \\
        - e^{ \left[ i\left( k_-+ \frac{\Delta q_\mathrm{pm}}{2} \right) L_\mathrm{a} \right]} 
        \frac{i g \bar{b}}{ \sqrt{v_1v_2} } \sin{\left[ \beta L_\mathrm{a} \right]  } &
        e^{ \left[ i\left( k_-+ \frac{\Delta q_\mathrm{pm}}{2} \right) L_\mathrm{a} \right]} 
        \left\{ \cos{\left[ \beta L_\mathrm{a} \right] } - \frac{i \Delta q_\mathrm{pm}}{2 \beta}  \sin{\left[ \beta L_\mathrm{a} \right] } \right\}
    \end{bmatrix}.
\end{multline}
In this equation, $\beta\equiv\sqrt{\left(\frac{\Delta q_\mathrm{pm}(\omega)}{2}\right) ^2 + \frac{\left| \bar{b}g \right|^2}{v_+ v_-}}$, with $v_{+(-)}$ denoting the group velocities of the symmetric and anti-symmetric modes in AOM region, and $\bar{b}$ and $g$ signifying the acoustic field and acoustic-optical coupling strength, respectively. 

To simplify the transfer matrix, we introduce the definitions: $ r(\omega)e^{i\phi_\mathrm{r}(\omega)} \equiv \cos{\left[ \beta L_\mathrm{a} \right] } + \frac{i \Delta q_\mathrm{pm}}{2 \beta}  \sin{\left[ \beta L_\mathrm{a} \right] }$,  $\mu(\omega) e^{i\phi_\mathrm{RF}} \equiv \frac{g\bar{b}}{\sqrt{v_1v_2}} \sin{\left[ \beta L_\mathrm{a} \right]}$, where $r(\omega), \mu(\omega)$ are real and $r(\omega)^2+\mu(\omega)^2=1$. $\phi_\mathrm{r}(\omega)$ can be understood as the phase originated from acoustic optical phase mismatch, and $\phi_{RF}$ is the acoustic phase that can be controlled by the driving RF phase.
\begin{equation}
    t_{a}^f(\phi_\mathrm{RF}) =\begin{bmatrix}
    r(\omega)e^{ \left[ i\left( k_+- \frac{\Delta q_\mathrm{pm}}{2} \right) L_\mathrm{a} \right]}e^{i\phi_\mathrm{r}(\omega)} &
        - i\mu(\omega) e^{ \left[ i\left( k_+- \frac{\Delta q_\mathrm{pm}}{2} \right) L_\mathrm{a} \right]} 
       e^{-i\phi_\mathrm{RF}} \\
        - i\mu(\omega) e^{ \left[ i\left( k_-+ \frac{\Delta q_\mathrm{pm}}{2} \right) L_\mathrm{a} \right]} 
        e^{i\phi_\mathrm{RF}}&
        r(\omega)e^{ \left[ i\left( k_-+ \frac{\Delta q_\mathrm{pm}}{2} \right) L_\mathrm{a} \right]}e^{-i\phi_\mathrm{r}(\omega)}
    \end{bmatrix}.
\end{equation}


The transfer matrix for the delay region is expressed as
\begin{equation}
    t_\mathrm{delay}  
    = \begin{bmatrix}
        e^{ik_\mathrm{d,+}(\omega_1)L_\mathrm{d,+}} & 0 \\
        0 & e^{ik_\mathrm{d,-}(\omega_2)L_\mathrm{d,-}}
    \end{bmatrix},
\end{equation}
where $k_\mathrm{d,+(-)}, L_\mathrm{d,+(-)}$ are the wavevector and delay length in delay region for symmetric mode and anti-symmetric mode.

For forward operation, given only input from $|+\rangle_{\mathrm{i},\omega_1}=a_0$, we can calculate the output as

\begin{equation}
    \begin{bmatrix} |+\rangle_{\mathrm{o},\omega_1} \\ |-\rangle_{\mathrm{o},\omega_2} \end{bmatrix}  = \\
    a_0\begin{bmatrix} r(\omega_1)^2e^{i[(2k_+-\Delta q_\mathrm{pm})L_\mathrm{a}+2\phi_\mathrm{r}+k_\mathrm{d,+}L_\mathrm{d,+}]}-\mu(\omega_1)^2e^{i[(k_++k_-)L_\mathrm{a}+k_\mathrm{d,-}L_\mathrm{d,-}+\phi_\mathrm{RF1}-\phi_\mathrm{RF2}]}\\ 
    -ir(\omega_1)\mu(\omega_1)(e^{i[(k_-+k_+)L_\mathrm{a}+\phi_\mathrm{RF2}+\phi_\mathrm{r}+k_\mathrm{d,+}L_\mathrm{d,+}]}+e^{i[(2k_-+\Delta q_\mathrm{pm})L_\mathrm{a}-\phi_\mathrm{r}+\phi_\mathrm{RF1}+k_\mathrm{d,-}L_\mathrm{d,-}]})\end{bmatrix}.
\end{equation}


Similarly, for backward operation, we have
\begin{equation}
    \begin{bmatrix} |+\rangle_{\mathrm{o},\omega_3} \\ |-\rangle_{\mathrm{o},\omega_2} \end{bmatrix}  
    = t_\mathrm{a}^b(\phi_\mathrm{RF1}) \cdot t_\mathrm{delay} \cdot t_\mathrm{a}^b(\phi_\mathrm{RF2}) 
    \begin{bmatrix} |+\rangle_{\mathrm{i},\omega_3} \\ |-\rangle_{\mathrm{i},\omega_2} \end{bmatrix}.
\end{equation}

The only difference between $t_a^f$ and $t_a^b$ is that $t_a^b$ has an extra phase mismatch term $q_\mathrm{pm}^b=q_\mathrm{pm}^f+2\frac{n_\mathrm{g,+}}{c}\Omega$. Since our interaction region is short, $2\frac{n_\mathrm{g,+}}{c}\Omega L_\mathrm{a}\ll1$, this term can be disregarded. Then we have $t_a^f \approx t_a^b$.

Given only input from $|-\rangle_{\mathrm{i},\omega_2}=a_0$, the output for backward operation can be calculated as
\begin{equation}
    \begin{bmatrix} |+\rangle_{\mathrm{o},\omega_3} \\ |-\rangle_{\mathrm{o},\omega_2} \end{bmatrix}  = \\
    a_0\begin{bmatrix} 
    -ir(\omega_3)\mu(\omega_3)(e^{i[(k_-+k_+)L_\mathrm{a}+\phi_\mathrm{RF1}+\phi_\mathrm{r}+k_\mathrm{d,-}L_\mathrm{d,-}]}+e^{i[(2k_++\Delta q_\mathrm{pm})L_\mathrm{a}-\phi_\mathrm{r}+\phi_\mathrm{RF2}+k_\mathrm{d,+}L_\mathrm{d,+}]})\\
    r(\omega_3)^2e^{i[(2k_--\Delta q_\mathrm{pm})L_\mathrm{a}+2\phi_\mathrm{r}+k_\mathrm{d,-}L_\mathrm{d,-}]}-\mu(\omega_3)^2e^{i[(k_++k_-)L_\mathrm{a}+k_\mathrm{d,+}L_\mathrm{d,+}+\phi_\mathrm{RF1}-\phi_\mathrm{RF2}]}
    \end{bmatrix}.
\end{equation}

For simplicity, we can assume small AOM conversion efficiency, $\mu(\omega)\ll1$, to study the frequency response of the system. With this assumption, we can write simplified output for forward operation as 
\begin{equation}
    \begin{bmatrix} |+\rangle_{\mathrm{o},\omega_1} \\ |-\rangle_{\mathrm{o},\omega_2} \end{bmatrix}  = 
    a_0\begin{bmatrix} 
    e^{i(2k_++k_\mathrm{d,+}L_\mathrm{d,+})}\\
    -ir(\omega_1)\mu(\omega_1)(e^{i[(k_-+k_++\Delta q_\mathrm{pm}/2)L_\mathrm{a}+\phi_\mathrm{RF2}+k_\mathrm{d,-}L_\mathrm{d,-}]}+e^{i[(2k_++\Delta q_\mathrm{pm}/2)L_\mathrm{a}+\phi_\mathrm{RF1}+k_\mathrm{d,+}L_\mathrm{d,+}]})
    \end{bmatrix}.
\end{equation}
Here we have used the fact that when $\mu(\omega)\ll1$, $2\phi_\mathrm{r}=\Delta q_\mathrm{pm}L_\mathrm{a}$.

Similarly, the simplified output for backward operation is
\begin{equation}
    \begin{bmatrix} |+\rangle_{\mathrm{o},\omega_3} \\ |-\rangle_{\mathrm{o},\omega_2} \end{bmatrix}  = 
    a_0\begin{bmatrix} 
    -ir(\omega_3)\mu(\omega_3)(e^{i[(k_-+k_++\Delta q_\mathrm{pm}/2)L_\mathrm{a}+\phi_\mathrm{RF1}+k_\mathrm{d,+}L_\mathrm{d,+}]}+e^{i[(2k_-+\Delta q_\mathrm{pm}/2)L_\mathrm{a}+\phi_\mathrm{RF2}+k_\mathrm{d,-}L_\mathrm{d,-}]})\\
    e^{i(2k_-+k_\mathrm{d,-}L_\mathrm{d,-})}
    \end{bmatrix}.
\end{equation}

In order to design an isolator with minimal insertion loss, we aim to maximize $|-\rangle_{\mathrm{o},\omega_2}$ in forward operation and simultaneously minimize $|+\rangle_{\mathrm{o},\omega_3}$ in backward operation. To achieve this, we derive the phase relations that satisfy these requirements
\begin{equation}
    (k_+(\omega_1)-k_-(\omega_2))L_\mathrm{a}+k_\mathrm{d,+}(\omega_1)L_\mathrm{d,+}-k_\mathrm{d,-}(\omega_2)L_\mathrm{d,-}+\phi_\mathrm{RF1}-\phi_\mathrm{RF2}=2n\pi,
\end{equation}
\begin{equation}
    (k_+(\omega_3)-k_-(\omega_2))L_\mathrm{a}+k_\mathrm{d,+}(\omega_3)L_\mathrm{d,+}-k_\mathrm{d,-}(\omega_2)L_\mathrm{d,-}+\phi_\mathrm{RF1}-\phi_\mathrm{RF2}=(2m-1)\pi.
\end{equation}
Here, $n$ and $m$ are integers. For our analysis, we consider the designed central optical frequency $\omega_0/2\pi$, with $\omega_1=\omega_0$, $\omega_3=\omega_2+\Omega=\omega_1+2\Omega$, and $\phi_\mathrm{RF1}=\phi_\mathrm{RF2}$. Using $ \text{(S.13)} - \text{(S.14)} $, we arrive at the minimum phase delay requirement (where $n = m$) as
\begin{equation}
    \frac{2\Omega n_\mathrm{g,+}L_\mathrm{a}}{c}+\frac{2\Omega n_\mathrm{g,+}^\mathrm{d}L_\mathrm{d,+}}{c} =\pi.
\end{equation}
This equation provides the design guidelines of the delay line length for two dispersion engineering methods, which we will discuss in more details in the following sections.

\section{Two-ridge delay method}

\subsection{Theory}

\begin{figure*}
    \includegraphics{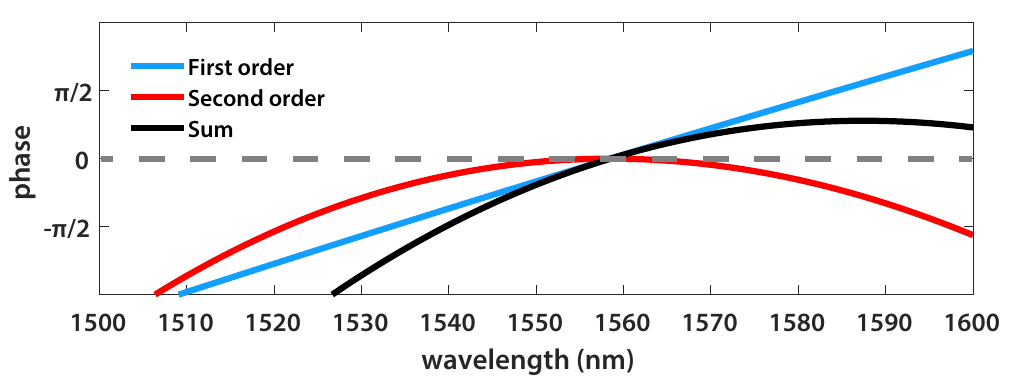}
    \caption{\label{figS3}
    Illustration of the contribution from different group delay terms using Device B fitted data. }
\end{figure*}

Here we adapt the theory derived in Section I.B to the dispersion engineered waveguide method. Following the equation S.13-14, the forward and backward can be viewed as the interference of two scattering terms. Then the anti-Stokes scattering output power in forward and backward operation can be written as
\begin{equation}
P_\mathrm{as}^{\mathrm{f}}\propto|a_\mathrm{as,0}|^2\cos^2{\frac{\phi_{\mathrm{f}(\omega)}}{2}};
\end{equation}
\begin{equation}
P_\mathrm{as}^{\mathrm{b}}\propto|a_\mathrm{as,0}|^2\cos^2{\frac{\phi_{\text{b}(\omega)}}{2}},
\end{equation}
where $|a_\mathrm{as,0}|^2=|g\bar{b}|^2 L_\mathrm{a}^2 \mathrm{sinc}^2(\Delta q_\mathrm{pm}(\omega)L_\mathrm{a}/2)$ and $\phi_{\text{f (b)}(\omega)}$ is phase difference between two terms for forward (backward) direction. In dispersion engineered waveguide method, we have $L_\mathrm{d,+}=L_\mathrm{d,-}=L_\mathrm{d}$, and using Taylor expansion, we have
\begin{equation}
\begin{split}
    \phi_{\text{f}}(\omega) = &(k_+(\omega) - k_-(\omega + \Omega))(L_\mathrm{a}+L_\mathrm{t}) + (k_+^\mathrm{d}(\omega) - k_-^\mathrm{d}(\omega + \Omega))L_\mathrm{d} + \phi_\mathrm{RF1}-\phi_\mathrm{RF2} \\
    &\quad= q(L_\mathrm{a}+L_\mathrm{t}) + \phi_\mathrm{RF1}-\phi_\mathrm{RF2} + \frac{n_\mathrm{g,+} - n_\mathrm{g,-}}{c}(\omega - \omega_0)(L_\mathrm{a}+L_\mathrm{t}) \\
    &+ \frac{n_\mathrm{g,+}^\mathrm{d}- n_\mathrm{g,-}^\mathrm{d}}{c}(\omega - \omega_0)L_\mathrm{d} +\frac{\beta_\mathrm{d,+}-\beta_\mathrm{d,-}}{2}(\omega - \omega_0)^2L_\mathrm{d}
\end{split}
\end{equation}


\begin{equation}
\begin{split}
    \phi_{\text{b}}(\omega) = &(k_+(\omega+ \Omega) - k_-(\omega ))(L_\mathrm{a}+L_\mathrm{t}) + (k_+^\mathrm{d}(\omega+ \Omega) - k_-^\mathrm{d}(\omega ))L_\mathrm{d} + \phi_\mathrm{RF1}-\phi_\mathrm{RF2} \\
    &\quad= \phi_{\text{f}}(\omega)+\left(\frac{n_\mathrm{g,+} + n_\mathrm{g,-}}{c}(L_\mathrm{a}+L_\mathrm{t})+\frac{n_\mathrm{g,+}^\mathrm{d}+ n_\mathrm{g,-}^\mathrm{d}}{c}L_\mathrm{d}\right)\Omega
\end{split}
\end{equation}
where $n_\mathrm{g,+(-)}^\mathrm{d}, \beta_\mathrm{d,+(-)}$ are the group index and group velocity dispersion for symmetric (anti-symmetric) mode.  It's worth to mention that such dispersion engineered waveguide is able to balance $n_\mathrm{g,+}$ and $n_\mathrm{g,-}$, but for $\beta_\mathrm{d,+(-)}$, they are not guaranteed to be balanced. In such case we need to take this higher order term into consideration. Here for device B we design second order term, $\frac{\beta_\mathrm{d,+}-\beta_\mathrm{d,-}}{2}(\omega - \omega_0)^2L_\mathrm{d}$, to be cancelling with first order terms at the wavelength we're interested, such that we have flat response and ultra wide-band isolation (Fig.~\ref{figS3}). And we also introduce $L_\mathrm{t}$ to account for the transition between the AOM and delay region.
For Fig.~3(c), the parameters used in the fitting are $n_\mathrm{g,+}^\mathrm{d}=3.5729, n_\mathrm{g,-}^\mathrm{d}=3.5721, n_\mathrm{g,+}=3.805, n_\mathrm{g,-}=3.885, \omega_0=\frac{2\pi c}{1540.65~\text{nm}}, L_\mathrm{a}=276~\text{um}, L_\mathrm{t}=274~\text{um}, L_\mathrm{d}=7.64~\text{mm}, \beta_\mathrm{d,+}-\beta_\mathrm{d,-}=-470\text{fs}^2/\text{mm}, \phi_\mathrm{RF1}-\phi_\mathrm{RF2}=-q(L_\mathrm{a}+L_\mathrm{t})$.

For Fig.~3(d), the parameters used in the fitting are $n_\mathrm{g,+}^\mathrm{d}=3.6847, n_\mathrm{g,-}^\mathrm{d}=3.6803, n_\mathrm{g,+}=3.8628, n_\mathrm{g,-}=3.9672, \omega_0=\frac{2\pi c}{1558.48~\text{nm}}, L_\mathrm{a}=276~\text{um}, L_\mathrm{t}=274~\text{um}, L_\mathrm{d}=7.64~\text{mm}, \beta_\mathrm{d,+}-\beta_\mathrm{d,-}=-372~\text{fs}^2/\text{mm}, \phi_\mathrm{RF1}-\phi_\mathrm{RF2}=-q(L_\mathrm{a}+L_\mathrm{t})$. Due to higher conversion efficiency in Device A, the simplified theory doesn't fit perfectly. For device B, the fitting agrees better to the experiment.

\subsection{Waveguide dispersion engineering}
To maintain the same group delay for the two spatial modes of the waveguide, we must engineer the waveguide's dispersion to assure identical group velocity for both modes. A significant challenge is that the geometry of the waveguide's cross-section simultaneously impacts the group delay for both spatial modes, making dispersion fine-tuning for individual modes rather intricate.

\begin{figure*}
    \includegraphics{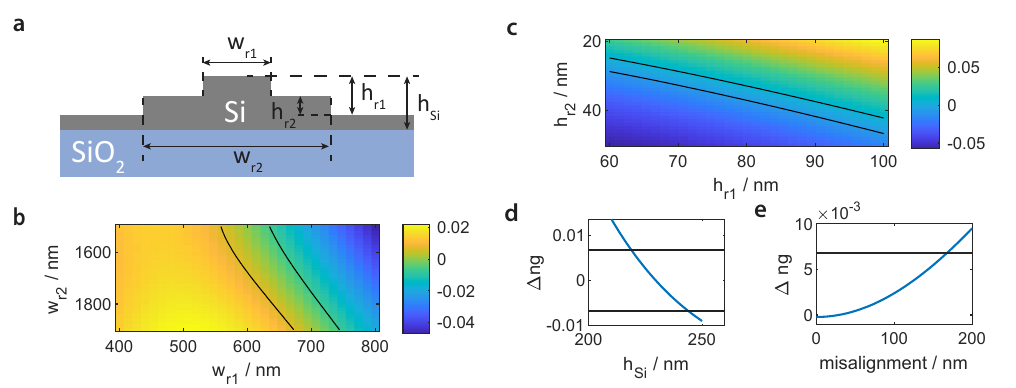}
    \caption{\label{figS4}
    \textbf{a}, Schematic of dispersion engineered two-ridge waveguide.
    \textbf{b,c}, Simulation of group index difference vs ridge widths and ridge thicknesses.
    \textbf{d,e}, Simulation of group index difference vs Si thickness and misalignment for two ridges.
    \textbf{b-e}, The black line indicates the requirement for 10nm 10dB isolation bandwidth.
    }
\end{figure*}

To overcome this obstacle, we introduce an innovative design incorporating an additional ridge to the conventional ridge waveguide, depicted in Fig.\ref{figS4}(a). The anti-symmetric mode is primarily defined by the wider ridge, while the symmetric mode is mainly determined by the narrower ridge. This allows for independent tuning of the dispersion properties by adjusting the geometry of the two ridges separately.

Fig.\ref{figS4}(b-c) represents the simulation of the group velocity difference for the two modes with various geometric parameters. The black line signifies the requirement to attain a 10nm bandwidth of a 10dB isolation ratio. It is noteworthy that the dispersion property is robust to waveguide width variation but is sensitive to etch thickness (5nm for ridge 2 and 10nm for ridge 1). Refining the etching recipe — such as employing a slower etching rate or atomic layer etching\cite{athavale1996realization} — could assist in accurately defining the ridge thickness.

To counteract fabrication imperfections and the disruption on dispersion from the active single ridge waveguide and the transition region, we target different etch thicknesses for three distinct wafers, as demonstrated in Table\ref{tab:wafer}.

However, we were only able to execute a liquid HF release on wafer II, constraining our bandwidth for device A. 
Through the simple combination of a liquid HF release and precise etch, we can achieve both low insertion loss and broad bandwidth in the same device.

\begin{table*}
    \caption{\label{tab:wafer}
    \textbf{Wafer etch thickness and bandwidth.} }
    \begin{ruledtabular}
    \begin{tabular}{lccccccr}
    Wafer number&
    \thead{Ridge 1 target}&
    \thead{Ridge 1 height}&
    \thead{Ridge 2 target}&
    \thead{Ridge 2 height}&
    \thead{10dB isolation ratio bandwidth}&
    \\
    \colrule
    Wafer I        &80nm&    $80.99\pm1.21$nm &33nm&   $30.13\pm0.35$nm    &  16nm    \\
    Wafer II       &80nm&    $74.73\pm0.73$nm &35nm&   $29.01\pm0.20$nm    &  5.6nm   \\
    Wafer III      &80nm&    $81.13\pm1.25$nm &37nm&   $31.78\pm0.36$nm    &  3.9nm   \\
    \end{tabular}
    \end{ruledtabular}
\end{table*}

\subsection{Loss analysis}

\begin{figure*}
    \includegraphics{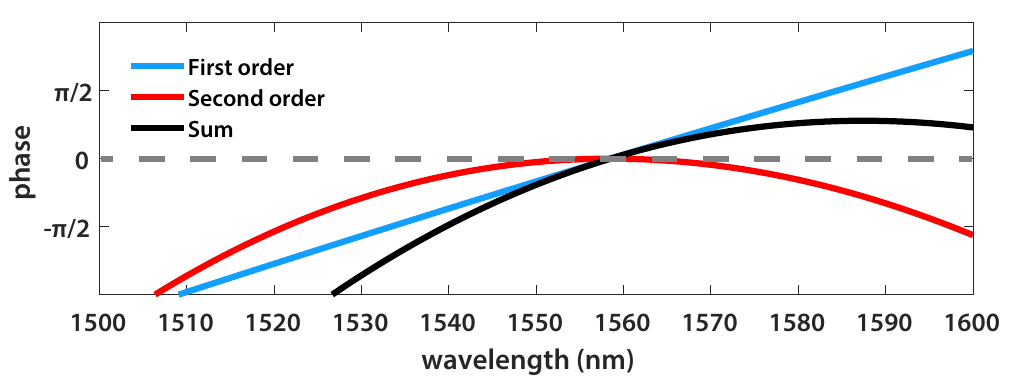}
    \caption{\label{figS5}
    \textbf{a}, Stitched microscope image of two ridge waveguide ring resonator.
    \textbf{b,c}, Drop port ring resonance response for symmetric and anti-symmetric mode respectively.
    \textbf{d,e}, Through and drop port ring single resonance response for symmetric and anti-symmetric mode respectively.
    \textbf{f}, Group index difference dependence on wavelength. Red dots are measurements from ring FSR, blue line is the parameter used in the fitting of Fig.2f}
\end{figure*}

To characterize the loss and dispersion of the two ridge waveguide, we utilize a racetrack ring resonator, as depicted in Fig.\ref{figS5}(a). The single point coupler can couple the symmetric mode and anti-symmetric mode to the ring resonator based on the input. It's essential to note that as the anti-symmetric mode is less confined, its coupling rate to the ring is much higher than the symmetric mode.

Fig.\ref{figS5}(b-c) illustrates the ring resonance response from the drop port measurement on the vapor HF wafer. By measuring the free spectral range (FSR) $\Delta\nu=\frac{c}{n_\mathrm{g}L}$ at different wavelengths, we can determine the group index difference for the two spatial modes.

To estimate the loss for the two-ridge waveguide, we inspect a single resonance. According to the coupled mode theory, the through and drop port response can be written as:

\begin{equation}
P_{\text{through}}=|1-\frac{2\kappa_e}{i(\omega-\omega_r)+\gamma+2\kappa_e}|^2
\end{equation}
\begin{equation}
P_{\text{drop}}=|\frac{-2\kappa_e}{i(\omega-\omega_r)+\gamma+2\kappa_e}|^2
\end{equation}

Here, $\kappa_e$ is the external coupling rate, and $\gamma$ is the intrinsic loss rate. From Fig.\ref{figS5}(d-e), we can extract the loss for the two-ridge waveguide that has undergone a vapor HF release to be $\alpha_{+} = 0.6661\pm0.0353\text{dB/cm}, \alpha_{-} = 0.2395\pm0.0051\text{dB/cm}$. We also measured the loss for the two-ridge waveguide that underwent a liquid HF release, yielding $\alpha_{+} = 0.3062\pm0.0023\text{dB/cm}, \alpha_{-} = 0.1864\pm0.0036\text{dB/cm}$.

To precisely characterize the insertion loss, we also measured the transition loss $L_{T\pm}$ from the active waveguide to the two-ridge waveguide, finding $L_{T-}=0.65\text{dB}$ for the anti-symmetric mode and negligible for the symmetric mode. The overall insertion loss for the isolator can be expressed as:

\begin{equation}
IL=E_{AO}+\frac{\alpha_++\alpha_-}{2}L_\mathrm{d}+L_{T-}+L_{MM-}
\end{equation}

Here, $E_{AO}$ denotes the conversion efficiency for the acoustic-optical scattering in dB, and $L_{MM-}$ represents the mode multiplexer loss for the anti-symmetric mode, as detailed in \cite{zhou2022intermodal}.

To measure insertion loss precisely, we measure the transmission for symmetric to anti-symmetric:

\begin{equation}
S2AS=2GL+L_{T-}+L_{MM-}+E_{AO}+\frac{\alpha_++\alpha_-}{2}L_\mathrm{d}
\end{equation}
and the transmission for symmetric to symmetric:

\begin{equation}
S2S=2GL+(1-E_{AO})+\alpha_+L_\mathrm{d}
\end{equation}
where $GL$ is the grating coupler loss.

Based on these data, we can compute $E_{AO}$ and, combined with $L_{MM-}, L_{T-}, \alpha_+, \alpha_-$, we have measured, we can calculate insertion loss accurately. The calculated insertion loss will provide a reliable measure of the efficiency of our two-ridge waveguide, thereby informing further optimization efforts.

\section{Two-waveguide delay method}

\subsection{Theory}

The primary difference between the two-waveguide and two-ridge approaches lies in the phase delay region. In the two-waveguide method, the phase delay can be expressed as follows:

For the forward direction, we have
\begin{equation}
\begin{split}
\phi_{\text{f}}(\omega) = &(k_+^\mathrm{a}(\omega) - k_-^\mathrm{a}(\omega + \Omega))L_\mathrm{a} + (k^\mathrm{d}(\omega)L_\mathrm{d,+} - k^\mathrm{d}(\omega + \Omega)L_\mathrm{d,-}) + \phi_\mathrm{RF1}-\phi_\mathrm{RF2} \\
&\quad= q(L_\mathrm{a}+L_\mathrm{t}) + \phi_\mathrm{RF1}-\phi_\mathrm{RF2} + \frac{n_\mathrm{g,+} - n_\mathrm{g,-}}{c}(\omega - \omega_0)L_\mathrm{a} + \frac{n_\mathrm{g}^\mathrm{d}L_\mathrm{d,+} - n_\mathrm{g}^\mathrm{d}L_\mathrm{d,-}}{c}(\omega - \omega_0)\\
&\quad+\frac{\beta_\mathrm{d}}{2}(\omega - \omega_0)^2(L_\mathrm{d,+}-L_\mathrm{d,-}). 
\end{split}
\end{equation}
And for the backward direction, we have
\begin{equation}
\begin{split}
\phi_{\text{b}}(\omega) = &(k_+^\mathrm{a}(\omega+ \Omega) - k_-^\mathrm{a}(\omega ))L_\mathrm{a} + (k^\mathrm{d}(\omega+ \Omega)L_\mathrm{d,+} - k^\mathrm{d}(\omega )L_\mathrm{d,-}) + \phi_\mathrm{RF1}-\phi_\mathrm{RF2} \\
&\quad= \phi_{\text{f}}(\omega)+\left(\frac{n_\mathrm{g,+} + n_\mathrm{g,-}}{c}L_\mathrm{a}+\frac{n_\mathrm{g}^\mathrm{d}L_\mathrm{d,+} + n_\mathrm{g}^\mathrm{d}L_\mathrm{d,-}}{c}\right)\Omega.
\end{split}
\end{equation}
Here, $k_\mathrm{d}(\omega)$, $n_\mathrm{g}^\mathrm{d}$  and $\beta_\mathrm{d}$ denote the wavevector, the group index and the group velocity dispersion of the single-mode waveguide in the phase delay region, respectively. In contrast to dispersion engineering, tuning waveguide length can simultaneously balance group delay difference and higher order delay originated from group velocity dispersion. In this case we can neglect this second order effect of dispersion properties, making our design much simpler. The parameters utilized in Fig.4(c) are $n_\mathrm{g}^\mathrm{d}L_\mathrm{d,+}= 30.618~\text{mm}, n_\mathrm{g}^\mathrm{d}L_\mathrm{d,-}= 30.578~\text{mm}, n_\mathrm{g,+}=3.8628, n_\mathrm{g,-}=3.9672, \omega_0=\frac{2\pi c}{1508.6~\text{nm}}, L_\mathrm{a}=270.8~\text{um}, \theta_{i}+qL_\mathrm{a}=7.56^\circ, \theta_{ii}+qL_\mathrm{a}=17.28^\circ, \theta_{iii}+qL_\mathrm{a}=39.96^\circ, \theta_{iv}+qL_\mathrm{a}=218.87^\circ$. The relative phase $\theta=\phi_\mathrm{RF1}-\phi_\mathrm{RF2}$ for i-iv aligns well with our experimental observations (Fig.4(c, i-iv)).

In practical scenarios, we cannot always assume that $L_\mathrm{d,+}$ is significantly greater than $L_\mathrm{a}$. 
Ideally, our goal is to satisfy the following equation:
\begin{equation}
\left(n_\mathrm{g,+} - n_\mathrm{g,-}\right) L_\mathrm{a} + n_\mathrm{g}^\mathrm{d} \left(L_\mathrm{d,+} - L_\mathrm{d,-}\right) = 0,
\end{equation}
which would allow for the expansion of the isolation bandwidth to infinity. 
Achieving this condition can be straightforward in the two-waveguide method, by adjusting the length difference between the upper and lower single-mode waveguides such that $L_\mathrm{d,+} - L_\mathrm{d,-} = -\frac{n_\mathrm{g,+} - n_\mathrm{g,-}}{n_\mathrm{g}} L_\mathrm{a} $. 

\subsection{Comparison with two-ridge method}
\begin{figure*}
    \includegraphics{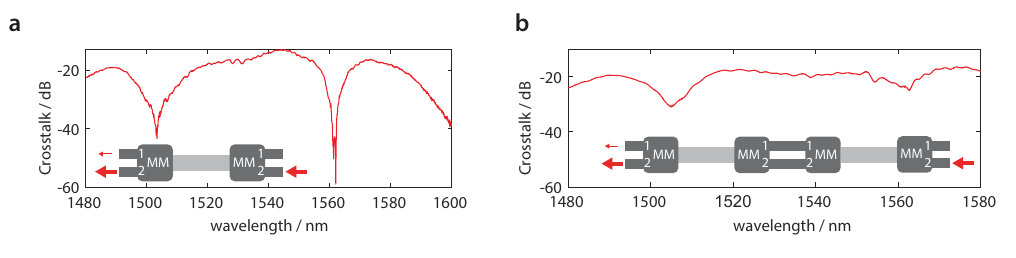}
    \caption{\label{figS6}
    \textbf{a}, Cross-talk measurement for two-ridge isolator.
    \textbf{b}, Cross-talk measurement for two-waveguide isolator. }
\end{figure*}

Two-waveguide method provides way of balanced phase delay robust to fabrication errors and without the need to consider group velocity dispersion. However, it will also requires additional two mode multiplexers. Since our mode multiplexer is based on directional coupler, it exhibits strong wavelength dependence, and only at certain narrow wavelength window we can have crosstalk less than -40dB. And this window varies from device to device. Adding additional 2 mode multiplexer will make the minimum crosstalk point hard to align, which increases the crosstalk floor and limits the isolation ratio, as is shown in Fig.\ref{figS6}.

\begin{figure*}
    \includegraphics{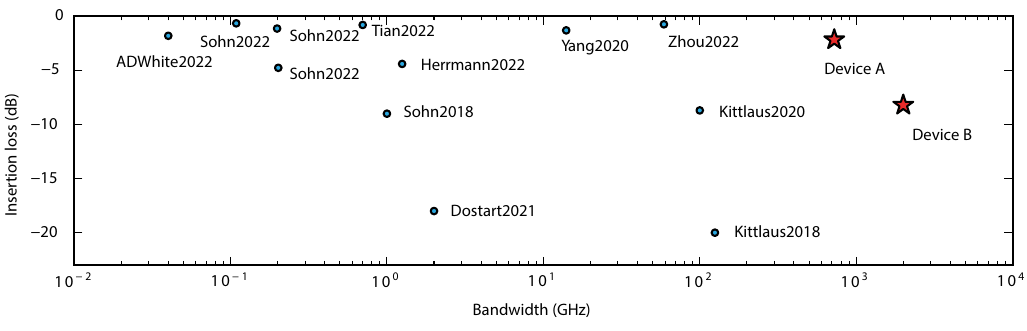}
    \caption{\label{figS7}
    \textbf{Review of on-chip non-magnetic optical isolators.} Collection of the previous results shown in \cite{Sohn2021,Tian2021,kittlaus2021electrically,sohn2018time,dostart2021optical,herrmann2022mirror,yang2020inverse,white2023integrated,zhou2022intermodal}.}
\end{figure*}

\begin{table*}
    \caption{\label{tab:compar}
    \textbf{Comparison of two dispersion engineering methods.} }
    \begin{ruledtabular}
    \begin{tabular}{lccccccr}
    \thead{Methods}&
    \thead{GVD}&
    \thead{Wafer thickness}&
    \thead{Intermodal crosstalk\\ in mode multiplexers}&
    \thead{Optical loss in\\ mode multiplexers}
    \\
    \colrule
    Two-ridge delay        &    More affected  &      Sensitive    &   Less affected     &   Less affected  \\
    Two-waveguide delay    &    Much less affected    &      Robust       &   More affected     &   More affected  \\
    \end{tabular}
    \end{ruledtabular}
\end{table*}

\begin{table}
    \caption{\label{tab:table_parameters}
    \textbf{The device parameters.} 
    Notice that the source of the parameters is indicated in the footnote. 
    Also, the slashed column is either not measured, or not discussed in this work. 
    }
    \begin{ruledtabular}
    \begin{tabular}{lcccr}
    \thead{Device parameters}&
    \thead{Two ridge device A}&
    \thead{Two ridge device B}&
    \thead{Two waveguide device}
    \\
    \colrule
    IDT pitch (nm)\footnotemark[1]          & 1750      & 1750      & 1750      \\
    IDT angle (deg)\footnotemark[1]         & 8.59   & 8.77   & 7.90  \\
    Active region length $L_\mathrm{a}$ (\textmu m)\footnotemark[1] & 276 & 276 & 271\\
    Delay length $L_{\mathrm{d}}$ (mm)\footnotemark[1]     &7.64        & 7.64      & 8.17 for symmetric mode\\
     & & & 8.16 for anti-symmetric mode\\
    Phononic crystal pitch (nm)\footnotemark[1]    & 688   & 688   & 688     \\
    Phononic crystal hole radius (nm) \footnotemark[1]    & 264.88 & 264.88 & 264.88\\
    Acoustic resonance frequency (GHz)\footnotemark[2]& 2.31     & 2.31      & 2.34    \\
    RF power (dBm)\footnotemark[2] & 36.04       & 32.38    & 30.24             \\
    Insertion loss (dB)\footnotemark[2] & -2.16       & -8.09    & -12.29             \\
    10dB isolation bandwidth (nm)\footnotemark[2] & 5.6       & 16    & /             \\
    Isolation contrast (dB)\footnotemark[2] & 12.4       & 28.1    & /             \\
    $n_{\mathrm{g},+}$\footnotemark[3]      & 3.805      & 3.863    & 3.863\\
    $n_{\mathrm{g},-}$\footnotemark[3]      & 3.885        & 3.967    & 3.967 \\
    $n_{\mathrm{g},+}^\mathrm{d}$\footnotemark[3]      & 3.573        & 3.685    & 3.747 \\
    $n_{\mathrm{g},-}^\mathrm{d}$\footnotemark[3]      & 3.572        & 3.680    & 3.747 \\
    $\beta_\mathrm{d,+}-\beta_\mathrm{d,-}(\mathrm{fs^2/mm})$\footnotemark[3] & -470       & -370   & / \\
    \end{tabular}
    \end{ruledtabular}
    \footnotetext{Device designs.}
    \footnotetext{Experimental measurements.}
    \footnotetext{Fig.~3 c, d fitting.}
\end{table}


\section{Comparison to previous works}
Here we present a review of current on-chip non-magnetic isolators, and compare their performances with our devices in Fig.~\ref{figS7}.

\section{Experimental setup}
Fig.~2a in the main text presents the experimental setup we utilized for characterizing the ultrawideband isolator. 
The microwave signal from the signal generator (SG) is split into two AOMs, with each AOM independently driving a spatial mode beamsplitting process. 
A microwave phase shifter (PS) is employed to control the relative phase between these two acousto-optic scatterers, which will be discussed in detail regarding the wavelength tunability of the system. 
The operation of the isolator in both forward and backward directions is characterized using a 2$\times$2 optical switch, with the optical output being simultaneously analyzed using two techniques. 
A heterodyne experiment, which takes the beatnote with the local oscillator, is utilized to perform frequency-resolved measurements, while an optical power meter is employed to determine the total optical power at the output port.


\end{document}